# Hierarchical DSSC structures based on "single walled" TiO$_2$ nanotube arrays reach back-side illumination solar light conversion efficiency of 8 %

*Seulgi So, Imgon Hwang, and Patrik Schmuki **

Department of Materials Science, WW4-LKO, University of Erlangen-Nuremberg (FAU), Martensstrasse 7, 91058 Erlangen, Germany
E-mail: schmuki@ww.uni-erlangen.de

**Abstract**

In the present work we introduce a path to the controlled construction of DSSCs based on hierarchically structured single walled, self-organized $TiO_2$ layers. In a first step we describe a simple approach to selectively remove the inner detrimental shell of anodic $TiO_2$ nanotubes (NTs). This then allows controlled well-defined layer-by-layer decoration of these $TiO_2$-NT walls with $TiO_2$ nanoparticles (this in contrast to conventional $TiO_2$ nanotubes). We show that such defined multiple layered decoration can be optimized to build dye sensitized solar cells that (under back-side illumination conditions) can yield solar light conversion efficiencies in the range of 8 %. The beneficial effects observed can be ascribed to a combination of three factors : 1) improved electronic properties of the "single walled" tubes themselves, 2) a further improvement of the electronic properties by the defined $TiCl_4$ treatment, and 3) a higher specific dye loading that becomes possible for the layer-by-layer decorated single walled tubes.

**Introduction**

Over the past decades, dye-sensitized solar cells (DSSCs) have attracted a great deal of interest in research and technology due to their high industrial potential towards low cost and versatile solar conversion technologies.[1] In the classic Grätzel-type cells the photoanode consists of a compacted $TiO_2$ nanoparticle film coated with a monolayer of a suitable inorganic dye.[1] The dye acts as light absorber where electrons are excited from the dye's HOMO (highest occupied molecular orbital) to the LUMO (lowest unoccupied molecular orbital) level,[1d,f] and from there are injected into the conduction band of the $TiO_2$ scaffold. Electrons then travel to the back contact and into an electrochemical conversion circuit. High conversion efficiency is reached if light absorption is maximized, and recombination – mainly with oxidized dye and electrolyte – is minimized (i.e., if the electron transfer rate through the $TiO_2$ network is faster than the various recombination pathways).[2] In order to suppress some potential drawbacks of nanoparticle based networks (e.g. random walk[2d] of carriers with an accordingly long diffusion path and recombination at grain junctions), over the past decade, considerable efforts have been devoted to the use of 1D scaffolds such as nanorods, nanowires, or nanotubes instead of nanoparticle layers.[2b, 3] However, many 1D structures provide a specific surface area that is 2-3 times lower than comparable layers fabricated from nanoparticles – thus the dye-loading per volume element of the photoanode is significantly lower. To overcome this issue, i.e. to combine a high surface area and directional electron transport, a number of hierarchical $TiO_2$ structures have been described in the literature.[4] For example, Grätzel et al.[4c] used $TiO_2$-coated fluorine-doped tin oxide nano-forest-like photoanode film grown by pulsed laser deposition to achieve fast electron transport, and decorated it with dye sensitized $TiO_2$ - this hierarchical mesostructure provided an overall DSSC efficiency of 4.9 %.

Over the past few years, a most frequently investigated 1D architecture are $TiO_2$-nanotube layers that are grown from Ti metal sheets by a simple self-organizing electrochemical anodization process.[5] This approach has the inherent advantage that the tube layers can be used directly in a "back-side illumination" configuration − i.e. using the Ti metal substrate, where the tubes are grown from, directly as a back contact for the $TiO_2$ photoanode, as illustrated in the supporting information (SI) figure S1a. The term "back-side illumination" is used here to maintain the classification as used in nanoparticle cells (where front-side refers to "through the $TiO_2$ layer" and back-side to "through the Pt back contact and electrolyte"). These "back-side" configurations (in the particle case) always have a lower efficiency than front-side illumination configuration DSSC (because in a back-side configuration some light is absorbed by the Pt-coated FTO and the iodine electrolyte).[1h]

Most efforts to increase the efficiency of such $TiO_2$ nanotube based cells target either smaller diameter nanotubes, bamboo geometries, or a secondary modification of the tube layers using etching or particle decoration to reach a higher specific surface area.[6] An overview of literature results from DSSCs based on $TiO_2$ nanotubes used in a back-side illumination configuration is given in figure S1b. Included are various efforts to change the tube dimensions and modify the active light harvesting area. Up to now, $TiO_2$-nanotube layers that hold a record DSSC back-side illumination efficiency of 7.12 % are layers that after formation again were processed with a hydrothermal treatment[4b] that enhanced the tube wall roughness and thus the specific dye adsorption. For front-side illumination nanotube-based DSSCs the world-record efficiency is > 9 %, nevertheless the approach is based on a tedious lift-off of limited size and fragile membranes from the substrate that then are attached on FTO with a thin layer of $TiO_2$ nanoparticles.[3h,i] For a more realistic (scalable) approach usually thin layers of Ti are evaporated on FTO, then completely anodized to transparency, and finally used in a classic solar cell assembly; however, for such structures the best front-side efficiency is ≈ 7 %.[7]

These nanotubes as well as most anodic $TiO_2$ nanotube layers used in the literature were grown using the most common growth approach in an ethylene glycol (EG) / fluoride electrolyte. Nevertheless, all these EG based anodic tubes, due to the nature of the formation mechanism, consist of a two layer (inner and outer shell) structure, as well established in literature,[8] and as illustrated in figure 1a and figure S2. This double walled nanotube morphology is only clearly apparent in SEM for tubes after annealing, or directly in TEM close to the tube bottom. In double walled tubes, the comparably thick inner layer contains a high amount of impurities and affects the sintering behavior during thermal treatments of the tubes.[8] Moreover, it narrows the inner diameter of the tubes and constrains the possibilities to modify the inner tube wall (e.g. by a well-controlled layer-by-layer $TiCl_4$ treatment or other $TiO_2$ particle decoration as we describe below).

In the present work we show that these conventional TiO$_2$ nanotube layers (valid for all tubes grown in EG) can be stripped from their inner layer by a suitable annealing / etching sequence, and we demonstrate that the resulting single walled nanotubes (figure 1b) allow a defined layer-by-layer coating with TiO$_2$ nanoparticles (figure 1c). This enables the careful adjustment and optimization of a hierarchical geometry towards high efficiency DSSCs (this in contrast to attempts using classic TiO$_2$ nanotubes grown in any EG electrolyte (figure S3). As a result, in this work we obtain conversion efficiencies for solar light close to 8 % (7.82 % ±0.2) with N719 dye[1f] (Ru-based dye (cis-bis (isothiocyanato) bis (2,2- bipyridyl 4,4-dicarboxylato) ruthenium(II) bis-tetrabutylammonium)) which is up to now the highest value reported for any back-side illuminated TiO$_2$-nanotube based DSSC. This effect is not only due to an increase of the active area but also due to accelerated charge transport observed for TiCl$_4$ treated nanotube walls.

**Results and discussion**

TiO$_2$ nanotube layers were grown in an EG/ lactic acid electrolyte[9] to a length of 16 μm as described in ref.9a. This approach leads to well anchored TiO$_2$ nanotube layers on the Ti metal substrate (which is advantageous for DSSC applications).[9b] After anodic growth, the tube walls consist, as expected, of a double shell structure evident from figure 1a.

In order to remove the inner shell we developed a selective core removal treatment as described in the supporting information (experimental details). It is based on the finding that a combination of a mild annealing treatment (150 °C) followed by an optimized piranha etch, leads to the selective dissolution of the inner tube shell.

After removal of the inner shell (figure S2), the inner diameter of the tubes is widened from approx. 40 nm to 110 nm (figure 1), with an inner wall surface that is smooth and well defined, and due to the stripping of the inner shell the amount of carbon contaminants strongly drops (figure S5, 6). The SEM and TEM images in figure 1b and in figure S2 confirm that the inner shell of the nanotubes has been removed completely, i.e. also at the bottom of the nanotubes. These opened tube layers then were layer-by-layer decorated with nanoparticles using an approach based on $TiCl_4$-hydrolysis (as commonly used in particle solar cells[1d]) and as described in the SI. We used conditions where each particle layer adds a thickness increment of approx. 13 nm with individual $TiO_2$ particles of ≈ 3 nm in diameter. Figure 1c shows SEM images of single walled nanotube layers after the first and fourth layer of decoration. In every case the formed tube walls were well and uniformly decorated with $TiO_2$ nanoparticles on the outside and inside (figure S4).

Figure 2 shows the solar cell performance for an increasing number of sequential $TiCl_4$ treatments using single and double walled $TiO_2$ nanotubes. For the single walled tubes a maximum efficiency of $\eta$ = 7.82 % (±0.2) is reached after four times decoration with $TiO_2$ nanoparticle layers. Clearly, the $TiCl_4$ decoration not only improves the surface area but also improves the fill factor. The reason for the decay in the performance for more than 4 layers is likely that for an even higher $TiO_2$ nanoparticle loading the electrolyte penetration into the tubes starts being hampered (see figure 1c after 5 layers), due to a clogging of the inner tube channel. The significance of the removal of the inner tube shell becomes clear, if the layer-by-layer treatment is attempted for conventional nanotube samples (figure 2b) - in this case already after applying nominally 2 layers, a maximum efficiency of $\eta$ = 5.53 % (± 0.14) is reached and overall a well-defined decoration is not possible even for this second layer, due to a non-homogeneous deposition on the tube walls (see figure S3).

Figure 2c provides additionally IPCE spectral data for the best solar cell. From these spectral data one can see that due to the use of a back-side configuration the typical photocurrent losses below 500 nm occur due to absorption in the iodine electrolyte.[1h] Most important, however, is that the single walled tubes show over the entire wavelength region a higher IPCE magnitude, and after optimized layer-by-layer particle decoration a significant photocurrent response at higher wavelengths. (Such a difference in the absorption behavior can be ascribed to a different reflectivity of the different tube morphologies (see figure S7).

It is also remarkable that the plain single walled tubes, although having a lower dye loading, show a higher $J_{SC}$ and a higher efficiency than double walled tubes, and even after one layer of decoration the performance of the single walled tubes ($\eta > 6\ \%$) is clearly higher than any value obtained for the double walled tubes. This shows that the increased specific surface area for dye decoration cannot be the sole reason for the observed improved performance of DSSC. In order to evaluate a possible influence of the treatments on the electron transport properties, we carried out IMPS and open circuit voltage decay measurements.

Figure 3a shows electron transfer time constants of double walled and single walled nanotube layers obtained from IMPS measurements. Clearly, electron transport in single walled nanotubes is faster than in double walled nanotube samples, and it is remarkable that an additional $TiCl_4$ treatment reduces the transport times even more, i.e. the treatment not only increases the surface area for dye adsorption but it also leads to improved electronic transport properties of the oxide scaffold.

The open circuit voltage decay measurements show a longer electron life time in double walled tubes than in single walled, however after the $TiCl_4$ treatment even longer life times can be measured. This indicates that while transport is faster in single walled tubes (figure 3a), the de-coring treatment may lead to some recombination sites (defects) – the latter can be strongly suppressed by the $TiCl_4$ treatment. Fig. 3b shows that the $TiCl_4$ treatment of a double walled tubes also increases the life time, but it is still lower than for the single walled $TiCl_4$ treated sample. (Please note that this goes along with the main beneficial effect of single walled tubes that is the considerably faster electron transfer time figure 3(a))

The fact that the improvement in solar cells is indeed due to an improvement of the $TiO_2$ scaffold is supported by IMPS measurements of the non-dye sensitized tubes measured under alternating UV excitation in a classic photoelectrochemical cell (figure 3c). From figure 3c bare single walled tubes show even in an aqueous electrolyte in conventional photoelectrochemical configuration clearly faster electron transport than double walled tubes, and again the $TiCl_4$ treatment is found to strongly improve the transport times. Thus, all above results are in line with an interpretation that a considerable part of the improvement in $TiCl_4$ treated single walled nanotubes is due to enhanced electron transport properties, likely caused by passivation effect of defects by the $TiCl_4$ treatment.

**Conclusion**

Overall, in the present work we demonstrate selective removal of the lower quality inner oxide shell of anodic $TiO_2$ nanotubes using a simple chemical etching process. The resulting single walled tubes allow a well-defined layer-by-layer $TiO_2$ nanoparticle decoration of the tube walls using repetitive $TiCl_4$ treatments. These hierarchical $TiO_2$ nanotube structures can significantly enhance the solar cell efficiency in dye sensitized solar cells. Using an optimized nanoparticle decoration we reach for DSSCs, using back-side illumination, an efficiency close to 8 %. The beneficial effects observed can be ascribed to a combination of three factors : 1) improved electronic properties of the "single walled" tubes themselves, 2) a further improvement of the electronic properties by the defined $TiCl_4$ treatment, and 3) a higher specific dye loading that becomes possible for the layer-by-layer decorated single walled tubes.

Finally, it is noteworthy that the simple annealing/chemical etching procedure introduced in this work to de-core TiO$_2$ nanotubes from their detrimental inner shell for solar cells should be generally applicable also to other applications of TiO$_2$ nanotubes. In particular, together with controlled TiCl$_4$ treatments this should provide an approach to significantly improve the electronic properties for virtually any photoelectrochemical TiO$_2$ nanotube application.

**Experimental Section**

To grow TiO$_2$ nanotube layers we used titanium foils (0.125 mm thick, 99.6+% purity, Advent, England) that were degreased by sonication in acetone, ethanol and isopropanol, rinsed with deionized water, and then dried with a nitrogen jet. Anodization was carried out with a high-voltage potentiostat (Jaissle IMP 88 PC) at 120V in a two-electrode configuration with a counter electrode made of platinum gauze using an electrolyte composition of 1.5 M lactic acid (LA, DL-Lactic acid, ~90%, Fluka), 0.1 M ammonium fluoride (NH$_4$F) and 5 wt% deionized H$_2$O in ethylene glycol (99 vol%) held at a temperature of 60 °C (HAAKE F3 Thermostat) for 1m 30s.[8] The formed anodic nanotube layers from a first anodization were removed by ultra-sonication. In a second anodization, we used the same experimental conditions. To remove the inner oxide shell (that consists of a carbon containing lower quality oxide[7]), the samples were annealed at 150 °C in air with a heating and cooling rate of 30 °C/min during 1 h using a Rapid ThermalAnnealer (Jipelec JetFirst100). Samples then were dipped in a piranha solution (H$_2$SO$_4$:H$_2$O$_2$ = 3:1) for 6 min at 70 °C. After immersion to H$_2$O and EtOH, the samples were dried with a nitrogen jet. Etching for significantly longer than the optimized time, the remaining nanotube layer starts to slowly dissolve too. Also, extended exposure can make the TiO$_2$ nanotube layer easier to peel off from the Ti substrate.

In order to convert the TiO$_2$ nanotubes to anatase, the samples were annealed at 450 °C in air with a heating and cooling rate of 30 °C/min during 1 h using a Rapid Thermal Annealer.[6b]

For morphological characterization, a field-emission scanning electron microscope (FE-SEM, Hitachi SEM FE 4800) was used. The thickness of the nanotubes was measured from SEM cross-sections. Further morphological and structural characterization of the TiO$_2$ nanostructures was carried out with a TEM (Philips CM30 TEM/STEM) (figure S2, S4). Composition and chemical state information were obtained by X-ray photoelectron spectroscopy (XPS, PHI 5600, US) (figure S5) and by Energy dispersive X-ray (EDX) (figure S6). X-ray diffraction analysis (XRD, X'pert Philips PMD with a Panalytical X'celerator detector) using graphite monochromized CuKα radiation (Wavelength 1.54056 Å) was used for determining the crystal structure of the samples (all samples used here were fully converted to anatase, figure S9).

For dye-sensitization, Ru-based dye (cis-bis (isothiocyanato) bis (2,2- bipyridyl 4,4-dicarboxylato) ruthenium(II) bis-tetrabutylammonium) (D719, Everlight, Taiwan, same as called N719 dye) was used. Samples were dye-sensitized by immersing for 1 day in a 300 mM solution of the Ru-based dye in a mixture of acetonitrile and tert-butyl alcohol (volume ratio: 1:1). After dye-sensitization, the samples were rinsed with acetonitrile to remove non-chemisorbed dye. To evaluate the photovoltaic performance, the sensitized nanotubes were sandwiched together with a Pt coated fluorine-doped glass counter electrode (TCO22-15, Solaronix) using a polymer adhesive spacer (Surlyn, Dupont). Electrolyte (0.60 M BMIM-I, 0.03 M I$_2$, 0.10 M GTC in acetonitril/ valeronitril (85:15 vol.)/ SB-163, IoLiTec Inc, Germany) was introduced into the space between the sandwiched cells. Using back-side illumination, the current-voltage characteristics of the cells were measured under simulated AM 1.5 illumination provided by a solar simulator (300 W Xe with optical filter, Solarlight), applying an external bias to the cell and measuring the generated photocurrent with a Keithley model 2420 digital source meter. The active area was defined by the opening of black shadow film mask to be 0.2 cm$^2$.

Incident photon-to-current conversion efficiency (IPCE) measurements were performed with a 150 W Xe arc lamp (LOT-Oriel Instruments) with an Oriel Cornerstone 7400 1/8 m monochromator. The light intensity was measured with an optical power meter.

For TiCl$_4$ treatments we used 0.1 M aqueous solutions of TiCl$_4$ prepared under ice-cooled conditions. The TiO$_2$ nanotube layers were then treated at 70 °C for 30 min. Afterwards, the samples were washed with DI water and rinsed with ethanol to remove any excess TiCl$_4$, and finally dried in a nitrogen jet. After the treatment, TiO$_2$ nanotube samples were annealed again at 450 °C for 10 min to crystallize attached nanoparticles.

Dye desorption measurements of the dye sensitized TiO$_2$ layers were carried out by immersing the samples in 5 ml of 10 mM KOH for 30 min. The concentration of fully desorbed dye was measured spectroscopically (using a Lambda XLS UV/VIS spectrophotometer, PerkinElmer) at 520 nm and calculated amount of dye absorption on the TiO$_2$ nanotube layer using the Beer–Lambert law.[1g] Intensity modulated photovoltage and photocurrent spectroscopy (IMPS) measurements were carried out using modulated light (10 % modulation depth) from a high power green LED (λ = 530 nm) and UV (λ = 325 nm). The modulation frequency was controlled by a frequency response analyzer (FRA, Zahner IM6) and the photocurrent or photovoltage of the cell was measured using an electrochemical interface (Zahner IM6), and fed back into FRA for analysis. The light incident intensity on the cell was measured using a calibrated Si photodiode.

**Acknowledgements**

The authors would like to acknowledge ERC, DFG and the Erlangen DFG cluster of excellence for financial support.

**Figure 1.** Process to build optimized hierarchical TiO$_2$ nanotube structures for DSSCs. (a) SEM images of "as formed" tubes (showing typical double walled morphology). (b) Tubes after core removal process leaving only outer tube shell present. (c) Layer-by-layer decoration with TiO$_2$ nanoparticles.

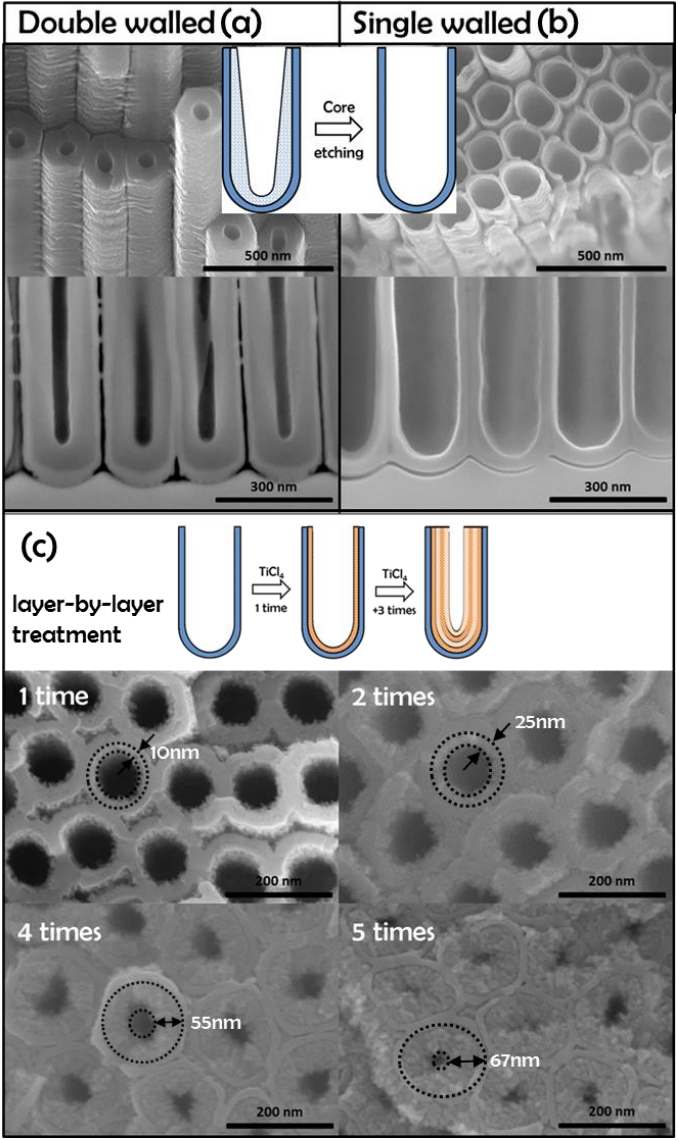

**Figure 2.** I–V characteristics for DSSCs fabricated using (a) single walled and (b) double walled TiO$_2$ nanotube samples without and with an increasing number of TiO$_2$ nanoparticle layers. (T(n) with n = number of added nanoparticle layers, J$_{SC}$ = short-circuit current, V$_{OC}$ = open-circuit voltage, *FF* = fill factor, η = efficiency). (c) IPCE spectra of the DSSCs.

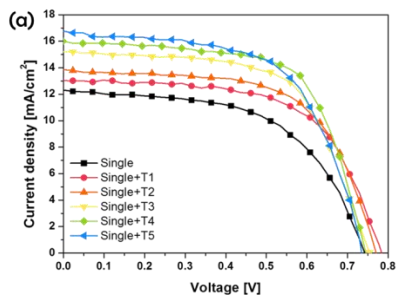

| Number of layers | V$_{OC}$ [V] | J$_{SC}$ [mA/cm$^2$] | FF [%] | η [%] | Dye absorption [nM/cm$^2$] |
|---|---|---|---|---|---|
| 0 | 0.74 | 12.31 | 56.43 | 5.14 | 102 |
| 1 | 0.78 | 13.02 | 61.05 | 6.20 | 137 |
| 2 | 0.77 | 13.86 | 61.94 | 6.61 | 170 |
| 3 | 0.76 | 15.20 | 62.50 | 7.22 | 211 |
| 4 | 0.75 | 16.01 | 65.12 | 7.82 | 244 |
| 5 | 0.73 | 16.75 | 60.52 | 7.40 | 269 |

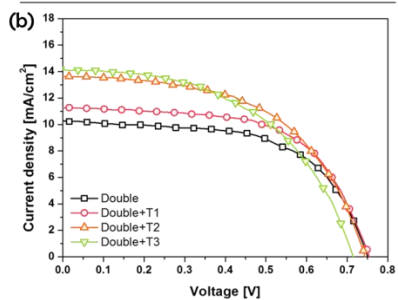

| | Number of layers | V$_{OC}$ [V] | J$_{SC}$ [mA/cm$^2$] | FF [%] | η [%] | Dye absorption [nM/cm$^2$] |
|---|---|---|---|---|---|---|
| Double walled | 0 | 0.75 | 10.28 | 58.84 | 4.54 | 116 |
| | 1 | 0.75 | 11.28 | 60.40 | 5.11 | 153 |
| | 2 | 0.74 | 13.65 | 54.75 | 5.53 | 200 |
| | 3 | 0.72 | 14.16 | 50.42 | 5.14 | 232 |

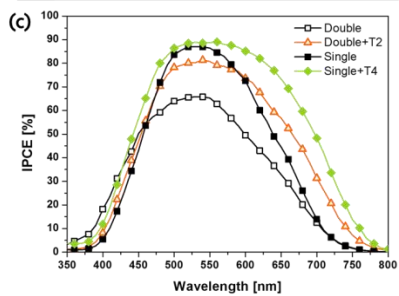

**Figure 3.** (a) Electron transfer time ($t_c$) constants from IMPS measurements for single walled and double walled $TiO_2$ nanotubes with T(n) layers of $TiO_2$ nanoparticle decoration (n = number of $TiO_2$ nanoparticle layers added). (b) The electron life time derived from open circuit voltage decay (insert) in corresponding DSSCs as a function of $V_{OC}$. (c) Electron transfer time ($t_c$) constants from IMPS measure\ements under the UV light (325nm) for single walled and double walled $TiO_2$ nanotubes with T(n) layers of $TiO_2$ nanoparticle decoration without dye adsorption in 0.1M $Na_2SO_4$ (in $H_2O$) electrolyte (n = number of $TiO_2$ nanoparticle layers added).

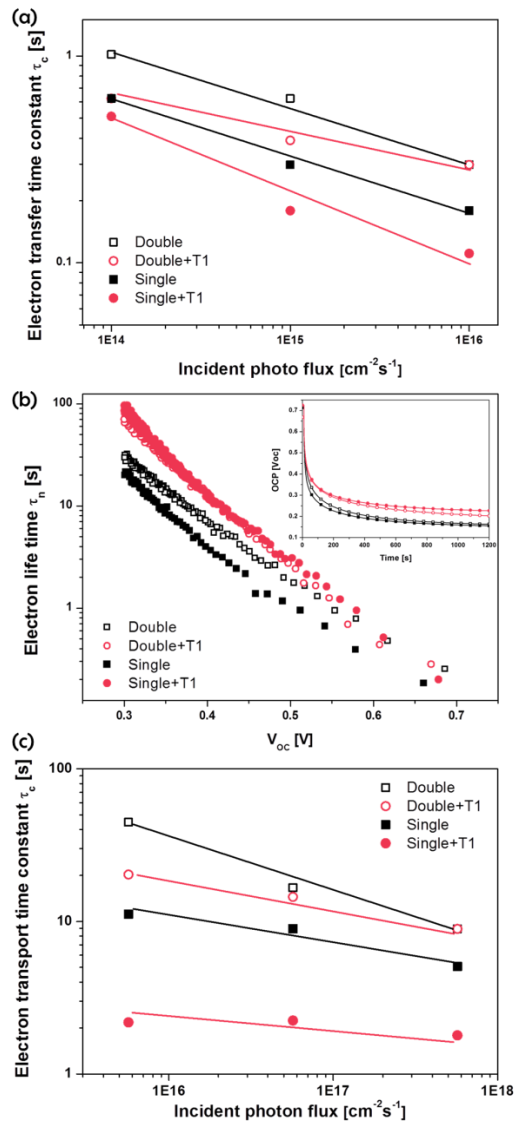

# Supporting Information

**Hierarchical DSSC structures based on "single walled" TiO$_2$ nanotube arrays reach back-side illumination solar light conversion efficiency of 8 %**

*Seulgi So, Imgon Hwang, and Patrik Schmuki\**

**Figure S1.**

(a) Schematic of a DSSC illustrated in a back-side configuration. (b) Overview of DSSC efficiencies in literature using TiO$_2$ nanotube layers in a back-side illumination configuration, compared with efficiency reached in the present work (red star); the insert shows schematically the de-coring / multiple layer decoration used in the present work.

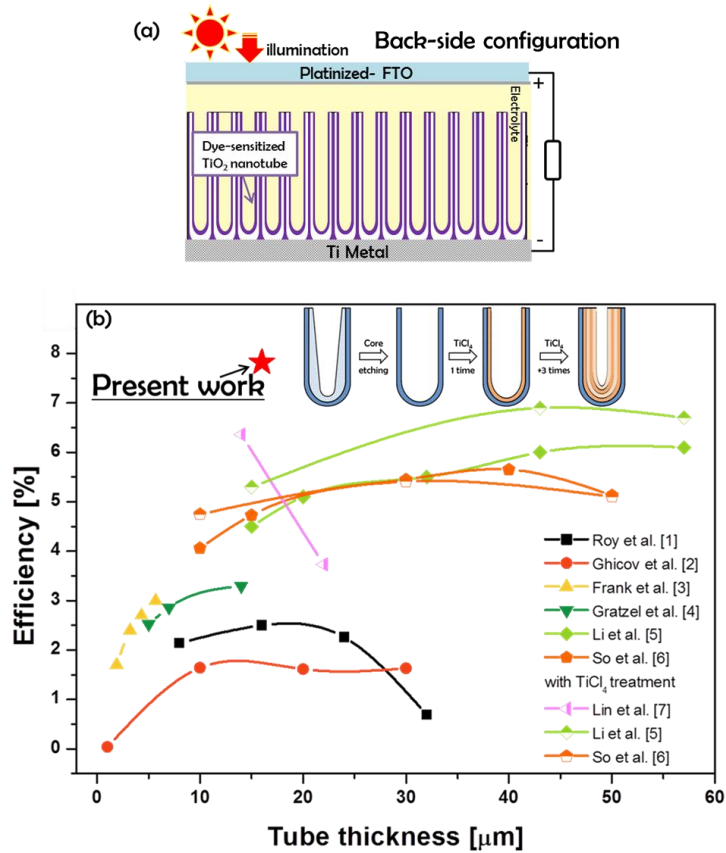

\* World-record efficiency of nanotube based front-side illumination DSSC is > 9 %, based on membrane transfer process on FTO with a thin layer of TiO$_2$ nanoparticle between the membrane and FTO.[8] For completely anodized thin film Ti on FTO approaches, the best front-side efficiency is 6.9 % [9].

**Figure S2**

SEM and TEM images of double and single walled nanotubes taken at the top surface and near the nanotube bottoms after annealing at 450 °C.

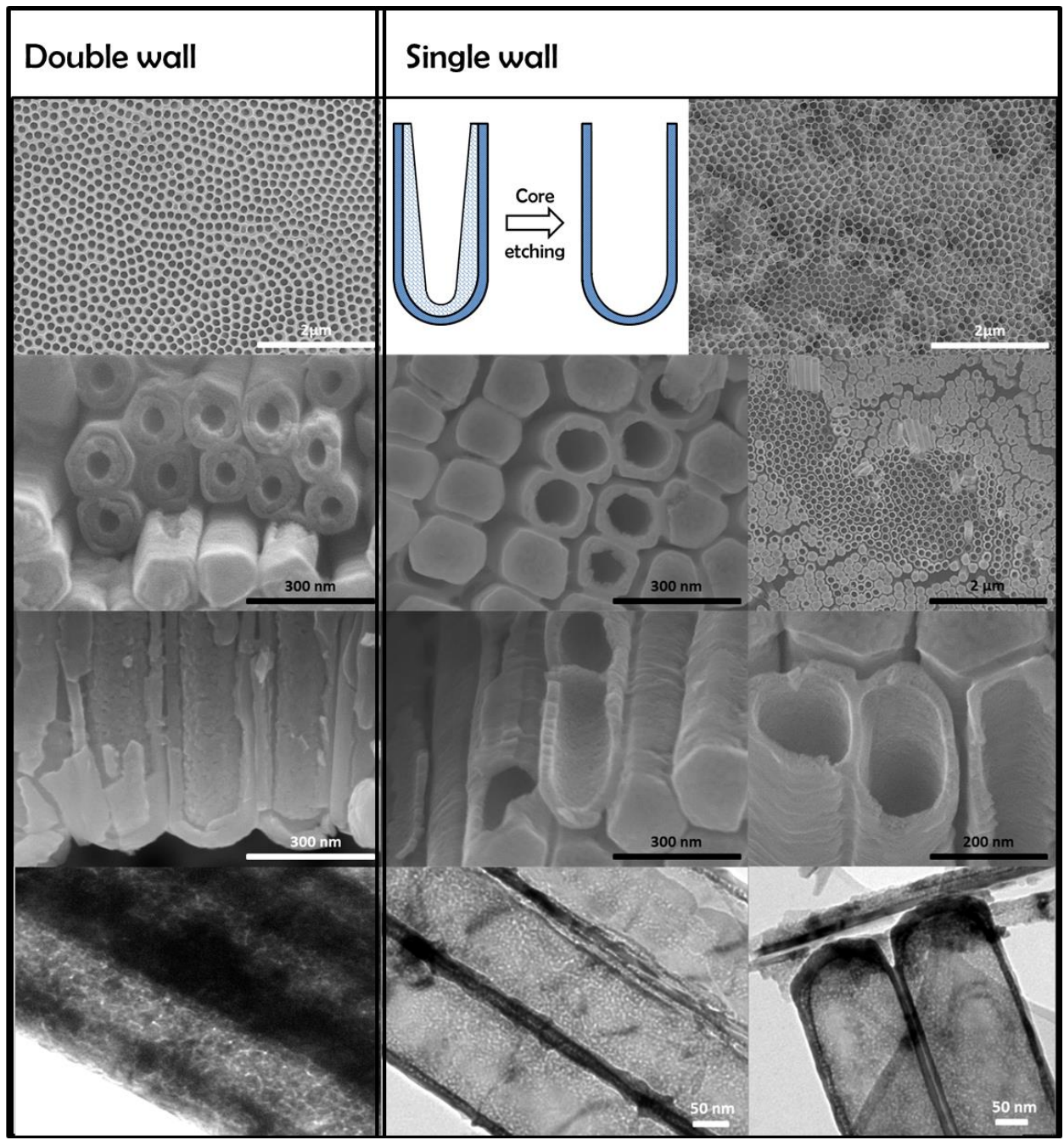

**Figure S3.**

SEM images double walled tubes after 2 times decoration with $TiO_2$ nanoparticles. This illustrates the incompatibility of classic tubes with a controlled layer-by-layer decoration (due to less defined inner shell of this tube type).

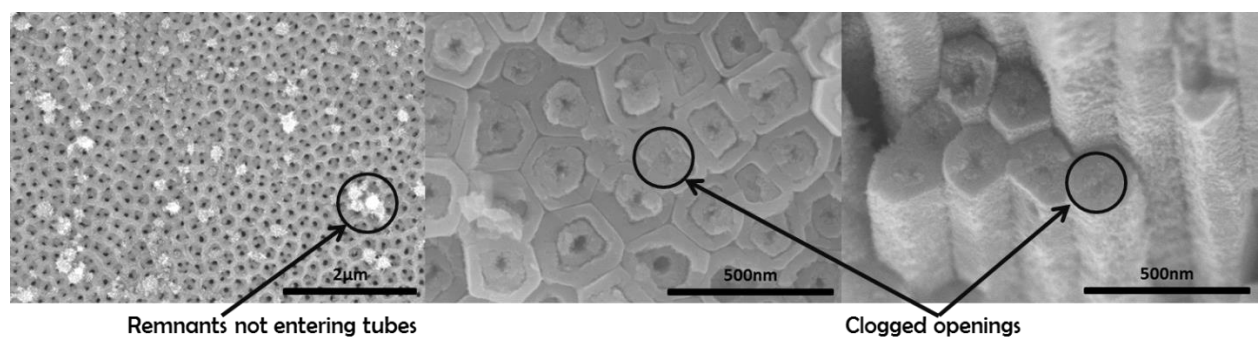

**Figure S4.**

SEM and TEM images of single walled nanotube layers after 1 time $TiCl_4$ treatment. Additionally at the bottom of nanotube layers also $TiO_2$ nanoparticles can be found.

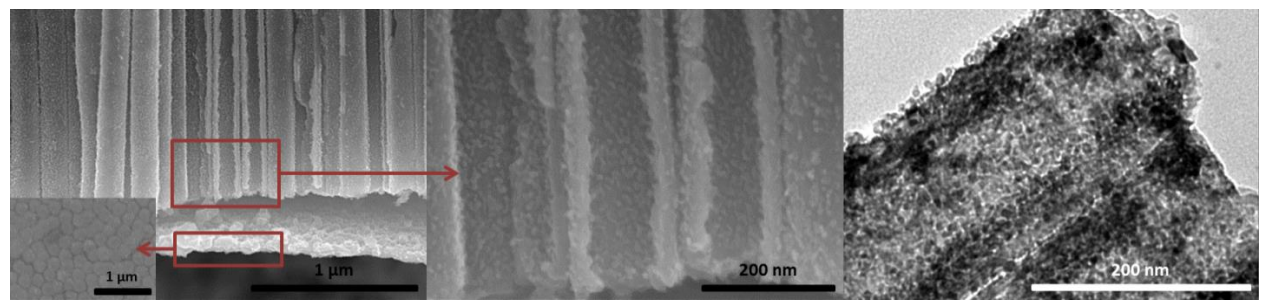

**Figure S5.**

XPS spectra taken on double walled and single walled nanotube samples after annealing at 150 °C and 450 °C. The results show that double walled nanotube samples have a high carbon content; after the core removal the single walled nanotube samples have a significantly lower carbon content.

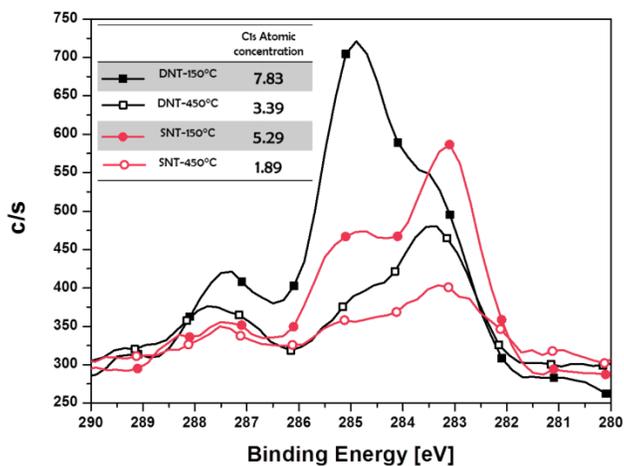

**Figure S6.**

EDX analysis for carbon taken at 3 different locations of cross sections of single- and double walled nanotubes.

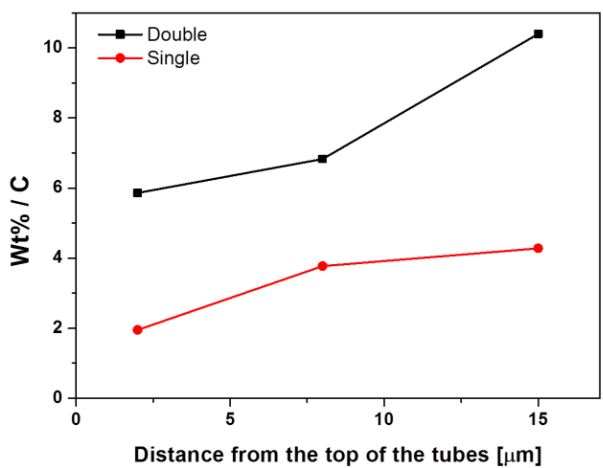

**Figure S7.** Diffuse reflectivity of the single- and double walled nanotubes with and without dye absorption.

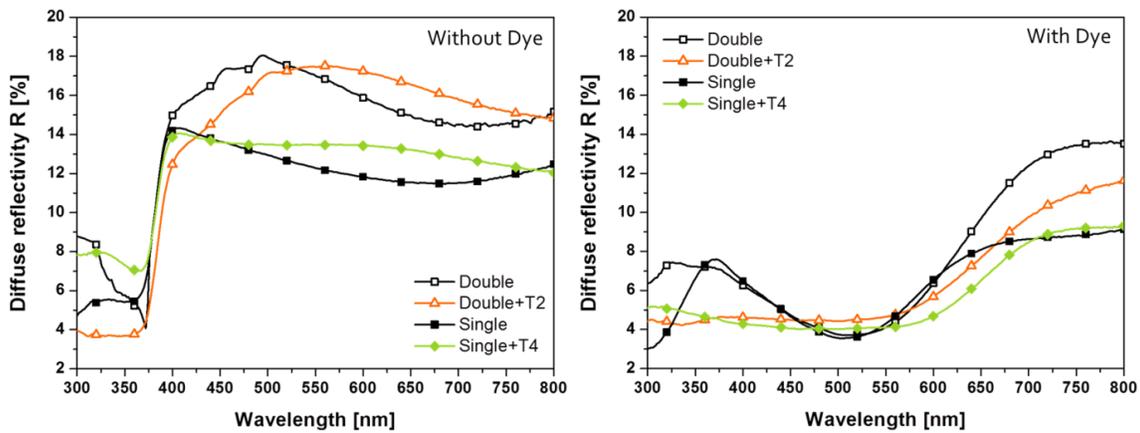

**Figure S8.** Comparison of (a) calculation are based on dye absorption using Beer lambert's law from the absorption spectra and (b) Kubelka-Munk value spectra calculated from diffuse reflectance measurements.

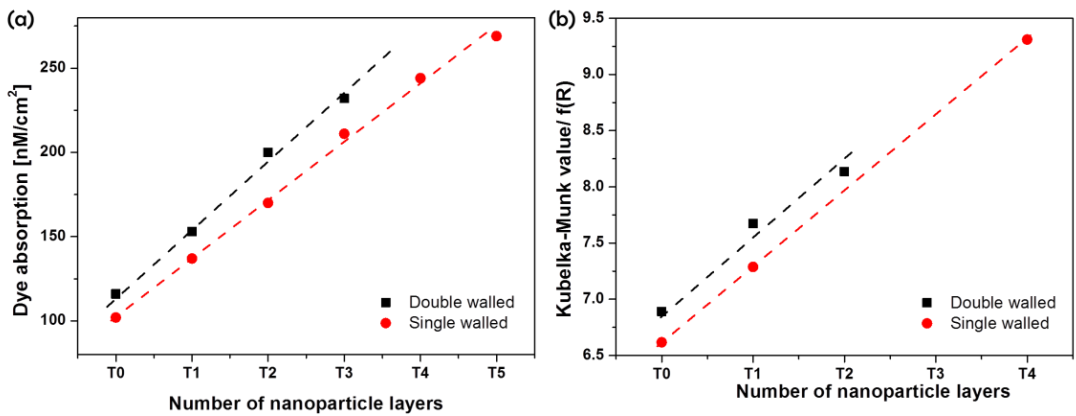

Here we compare dye loading by dissolving the dye from the $TiO_2$ nanotube samples using 10 mM KOH solution and then calculate dye absorption on the layers using the Beer Lambert's law (as frequently used in literature[10]) with measurements of the diffuse reflectance and convertion to absorption in the absorbance region of the dye (600nm). Both data show that double walled samples have higher amount of dye adsorbed. (This is because inner wall from the double walled nanotube after annealing gets rough and thus gives additional surface area for dye adsorption.) Both, single and double walled samples, when loading additional nanoparticle layers, also show an increase in the amount of dye adsorption, **but** the double walled $TiO_2$ nanotube layers (with more than 2 times nanoparticle decoration) start to clog openings and do not allow a defined layer by layer decoration. (see in figure S3)

**Figure S9.**

XRD spectra taken on double walled and single walled nanotube samples after annealing at 150 °C and 450 °C. The thermal treatment at 150 °C has no apparent effect on the crystallinity, after annealing at 450 °C conversion to anatase has occurred. (A = anatase, Ti = Titianium)

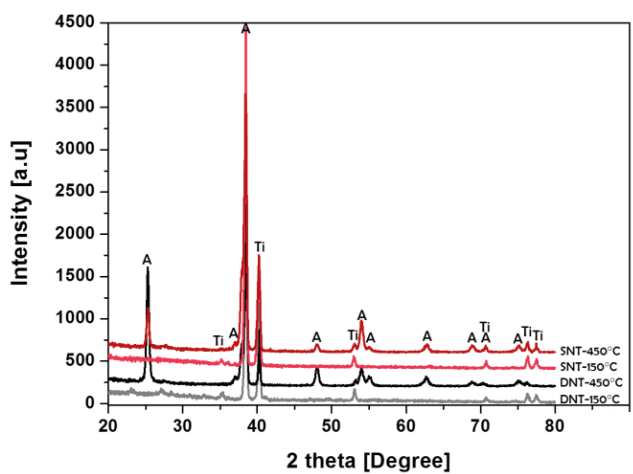